\documentstyle[aps]{revtex}

\begin{document}
\title{The direction of time: from the global arrow to the local arrow}
\author{Mario Castagnino}
\address{Instituto de Astronom\'\i a y F\'\i sica del Espacio\\
Casilla de Correos 67, Sucursal 28, 1428 Buenos Aires, Argentina}
\author{Luis Lara}
\address{Departamento de F\'\i sica, Universidad Nacional de Rosario\\
Av. Pellegrini 250, 2000 Rosario, Argentina}
\author{Olimpia Lombardi}
\address{CONICET - Universidad de Buenos Aires \\
Pu\'an 470, 1406 Buenos Aires, Argentina}
\maketitle

\begin{abstract}
In this paper we discuss the traditional approaches to the problem of the
arrow of time. On the basis of this discussion we adopt a global and
non-entropic approach, according to which the arrow of time has a global
origin and is an intrinsic, geometrical feature of space-time. Finally, we
show how the global arrow is translated into local terms as a local
time-asymmetric flux of energy.
\end{abstract}

\section{Introduction}

Since the nineteenth century, the problem of the direction of time has been
one of the most controversial questions in the foundations of physics. Many
theoretical contributions have been made in the seeking of an answer to the
problem. However, despite of all the debates, very little progress towards a
consensus has been achieved. Our impression is that this situation is mainly
due to the fact that different concepts are usually confused in the
discussions and different problems are traditionally subsumed under the same
label. For this reason, we will attempt to disentangle and clarify some of
the issues involved in the debates about the direction of time.

In particular, we will argue that it is necessary to carefully distinguish
between the problem of irreversibility and the problem of the arrow of time:
whereas the first one can be addressed in local terms, the second one
requires global considerations. On this basis, we will define the arrow of
time as an intrinsic, geometrical feature of space-time, rejecting the
traditional entropic approach according to which the direction of time is
defined by the gradient of the entropy function of the universe.

\section{Time-reversal invariance and irreversibility}

In general, both concepts are invoked in the treatment of the problem of the
arrow of time, but usually with no elucidation of their precise meanings;
this results in confusions that contaminate many interesting discussions.
For this reason, we will start from providing some necessary definitions.

Time-reversal invariance is a property of dynamical equations (laws) and, 
{\it a fortiori}, of the set of its solutions (evolutions). Reversibility is
a property of a single solution of a dynamical equation.

{\bf Def.: }A dynamical equation is {\it time-reversal invariant} if it is
invariant under the transformation $t\rightarrow -t$; as a result, for each
solution $f(t)$, $f(-t)$ is also a solution.

{\bf Def.:} A solution of a dynamical equation is {\it reversible} if it
corresponds to a closed curve in phase space.

It is quite clear that both concepts are different to the extent that they
apply to different entities: equations and solutions respectively.
Furthermore, they are not even correlated. In fact, time-reversal invariant
equations can have irreversible solutions. For instance, the equation of
motion of the pendulum with Hamiltonian:

\[
H=\frac 12\,p_\theta ^2+\frac{K^2}2\,\cos \theta 
\]
is time-reversal invariant, namely, it is invariant under the transformation 
$\theta \rightarrow \theta ,$ $p_\theta \rightarrow -p_\theta $; however,
whereas the trajectories within the separatrices are reversible since they
are closed curves, the trajectories above (below) the separatrices are
irreversible since, in the infinite time-limit, $\theta \rightarrow \infty $
($\theta \rightarrow -\infty )$.

When both concepts are elucidated in this way, {\it the problem of
irreversibility} can be clearly stated: {\it how to explain irreversible
evolutions in terms of time-reversal invariant laws}. But once it is
recognized that irreversibility and time-reversal invariance apply to
different entities, it is easy to find a conceptual answer to the problem:
nothing prevents a time-reversal invariant equation from having irreversible
solutions. Of course, this answer does not provide the full solution of the
problem: a great deal of theoretical work is needed for obtaining
irreversible evolutions from an underlying time-reversal invariant dynamics
(see, for instance, \cite{Irrev}). Here we only mean that, in order to face
the problem of irreversibility, the question about the arrow of time does
not need to be invoked: the distinction between the two directions of time
is usually assumed when the irreversible evolutions are conceived as
processes going from non-equilibrium to equilibrium or to preparation to
measurement towards the future.

\section{What is ''the problem of the arrow of time''?}

Traditional discussions around the problem of the arrow of time are usually
subsumed under the label ''the problem of the direction of time'', as if we
could find an exclusively physical criterion for singling out {\it the}
direction of time, identified with what we call ''the future''. But there is
nothing in local physics that distinguishes, in a non-arbitrary way, between
past and future. It might be objected that physics implicitly assumes this
distinction with the use of temporally asymmetric expressions, like ''future
light cone'', ''initial conditions'', ''increasing time'', and so on.
However this is not the case, and the reason relies on the distinction
between ''conventional'' and ''substantial''.

Two objects are {\it formally identical} when there is a permutation that
interchanges the objects but does not change the properties of the system to
which they belong. In physics it is usual to work with formally identical
objects: the two lobes of a light cone, the two spin senses, etc.

i.- We will say that we establish a {\it conventional} difference when we
call two formally identical objects with two different names, e.g{\it .},
when we assign different signs to the two spin senses.

ii.- We will say that the difference between two objects is {\it substantial}
when we give different names to two objects which are not formally identical
(see \cite{Penrose}, \cite{Sachs}). In this case, even though the names are
conventional, the difference is substantial. E.g., the difference between
the two poles of the theoretical model of a magnet is conventional since
both poles are formally identical; the difference between the two poles of
the Earth is substantial because in the north pole there is an ocean and in
the south pole there is a continent (and the difference between ocean and
continent remains substantial even if we conventionally change the names of
the poles).

Once this point is accepted, the problem cannot yet be posed in terms of
singling out the future direction of time: the problem of the arrow of time
becomes the problem of finding a {\it substantial difference} between the
two temporal directions. But if this is our central question, we cannot
project our independent intuitions about past and future for solving it
without begging the question. If we want to address the problem of the arrow
of time from a perspective purged of our temporal intuitions, we must avoid
the conclusions derived from subtly presupposing time-asymmetric notions. As
Huw Price \cite{Price} claims, it is necessary to stand at a point outside
of time, and thence to regard reality in atemporal terms: this is ''{\it the
view from nowhen}''. This atemporal standpoint prevents us from using the
temporally asymmetric expressions in a non-conventional way: the assumption
about the difference between past and future or between preparation and
measurement is not yet legitimate in the context of the problem of the arrow
of time.

But then, what does ''the arrow of time'' mean when we accept this
constraint? Of course, the traditional expression coined by Eddington has
only a metaphorical sense: its meaning must be understood by analogy. We
recognize the difference between the head and the tail of an arrow on the
basis of its geometrical properties; therefore, we can substantially
distinguish between both directions, head-to-tail and tail-to-head,
independently of our particular perspective. Analogously, we will conceive 
{\it the problem of the arrow of time} in terms of {\it the possibility of
establishing a substantial distinction between the two directions of time on
the basis of exclusively physical arguments}.

\section{Traditional approaches}

\subsection{The traditional local approach}

The traditional local approach owes its origin to the attempts of reducing
thermodynamics to statistical mechanics: in this context, the usual answer
to the problem of the arrow of time consists in defining the future as the
direction of time in which entropy increases. However, already in 1912 Paul
and Tatiana Ehrenfest \cite{Ehrenfest} noted that, if the entropy of a
closed system increases towards the future, such increase is matched by a
similar one in the past of the system. In other words, if we trace the
dynamical evolution of a non-equilibrium system at the initial time back
into the past, we will obtain states that are more uniform than the
non-equilibrium initial state. Gibbs' answer to the Ehrenfests' challenge
was based on the assumption that probabilities are determined from prior
events to subsequent events. But this answer clearly violates the ''nowhen''
standpoint: probabilities are blind to temporal direction; then, any
resource to the distinction between prior and subsequent events commits a 
{\it petitio principii} by presupposing the arrow of time from the start.

It is interesting to note that this old discussion can be generalized to the
case of any kind of irreversible evolution arising from time-reversal
invariant laws. In fact, time-reversal invariant equations always produce ''%
{\it t-symmetric twins}'', that is, two mathematical structures
symmetrically related by a time-reversal transformation: each ''twin'',
which usually represents an irreversible evolution, is the temporal mirror
image of the other ''twin''. For instance, electromagnetism provides a pair
of advanced and retarded solutions, that are usually related with incoming
and outgoing states in scattering situations as described by Lax-Phillips
scattering theory \cite{Lax-Phillips}. In irreversible quantum mechanics,
the analytical extension of the energy spectrum of the quantum system's
Hamiltonian into the complex plane leads to poles in the lower half-plane
(usually related with decaying unstable states), and symmetric poles in the
upper half-plane (usually related with growing unstable states) (see \cite
{Cast-Laura}). However, at this level the twins are only conventionally
different: we cannot distinguish between advanced and retarded solutions or
between lower and upper poles without assuming temporally asymmetric
notions, as the asymmetry between past and future or between preparation and
measurement. Here the real challenge consists in supplying a
non-conventional criterion for choosing one of the twins as the physically
relevant: such a criterion must establish a substantial difference between
the two members of the pair. But it is precisely this kind of criterion what
exceeds the context of local physics.

The problem can also be posed in different terms. Let us accept that we have
solved the irreversibility problem; so we have the description of all the
irreversible evolutions, say, decaying processes, of the universe. However,
since we have not yet established a substantial difference between both
directions of time, we have no way to decide towards which temporal
direction each decay proceeds. Of course, we would obtain the arrow of time
if we could coordinate the processes in such a way that all of them
parallelly decay towards the same temporal direction. But this is precisely
what local physics cannot offer: only by means of global considerations all
the decaying processes can be coordinated. This means that the global arrow
of time plays the role of the background scenario where we can meaningfully
speak of the temporal direction of irreversible processes, and this scenario
cannot be established by local theories that only describe phenomena
confined in small regions of space-time.

\subsection{The traditional global approach}

When, in the late nineteenth century, Boltzmann developed the probabilistic
version of his theory in response to the objections raised by Loschmidt and
Zermelo, he had to face a new challenge: how to explain he highly improbable
current state of our world. In order to answer this question, Boltzmann \cite
{Boltzmann} offered the first cosmological approach to the problem. Since
this seminal work, many authors have related the temporal direction
past-to-future to the gradient of the entropy function of the universe. For
instance, Feynman asserts: {\bf ''For some reason, the universe at one time
had a very low entropy for its energy content, and since then entropy has
increased. So that is the way towards future. That is the origin of all\
irreversibility''} \cite{Feynman}. In a similar sense, Davies claims that 
{\bf ''There exists an arrow of time only because the universe originates in
a less-than-maximum entropy state''} \cite{Davies}. Even if these authors
admit the need of global arguments for solving the problem of the arrow of
time, they coincide in considering that it must be addressed in terms of
entropy.

The global entropic approach rests on two assumptions: that it is possible
to define entropy for a complete cross-section of the universe, and that
there is an only time for the universe as a whole. However, both assumptions
involve difficulties. In the first place, the definition of entropy in
cosmology is still a very controversial issue: there is not a consensus
regarding how to define a global entropy for the universe. In fact, it is
usual to work only with the entropy associated with matter and radiation
because there is not yet a clear idea about how to define the entropy due to
the gravitational field. In the second place, when general relativity comes
into play, time cannot be conceived as a background parameter which, as in
pre-relativistic physics, is used to mark the evolution of the system.
Therefore, the problem of the arrow of time cannot legitimately be posed,
from the beginning, in terms of the entropy gradient between the two ends of
a linear and open time.

Nevertheless, these points are not the main difficulty: there is a
conceptual argument for abandoning the traditional entropic approach.
Entropy is a phenomenological property: a given value of entropy is
compatible with many configurations of a system. The question is whether
there is a more fundamental property of the universe which allows us to
distinguish between both temporal directions. On the other hand, if the
arrow of time reflects a substantial difference between both directions of
time, it is reasonable to think that it is an intrinsic property of time, or
better, of space-time, and not a secondary feature depending on a
phenomenological property. For these reasons we will follow Earman's ''{\it %
Time Direction Heresy}'' \cite{Earman}, according to which the arrow of time
is an intrinsic, geometrical property of space-time which does not need to
be reduced to a non-temporal feature as entropy. In other words, the
geometrical approach to the problem of the arrow of time has conceptual
priority over the entropic approach, since the geometrical properties of the
universe are more basic than its thermodynamic properties.

\section{Conditions for a global and non-entropic arrow of time}

\subsection{Temporal orientability}

In a Minkowski space-time, it is always possible to define the class of all
the future light semi-cones (lobes) and the class of all the past light
semi-cones (where the labels ''future'' and ''past'' are conventional). In
general relativity the metric can always be reduced, in small regions of
space-time, to the Minkowski form. However, on the large scale, we do not
expect the manifold to be flat because gravity can no longer be neglected.
Many different topologies are consistent with Einstein's field equations; in
particular, the possibility arises of space-time being curved along the
spatial dimension in such a way that the spacelike sections of the universe
become the three-dimensional analogous of a Moebius band; in technical terms
it is said that the space-time is temporally non-orientable.

{\bf Def.:} A space-time is {\it temporally orientable} if there exists a
continuous non-vanishing vector field on it which is timelike with respect
to its metric.

By means of this field, the set of all lobes of the manifold can be split
into two equivalence classes, $C_{+}$ and $C_{-}$: the lobes of $C_{+}$
contain the vectors of the field and the lobes of $C_{-}$ do not contain
them. On the other hand, in a temporally non-orientable space-time it is
possible to transform a future pointing timelike vector into a past pointing
timelike vector by means of a continuous transformation that always keeps
non-vanishing timelike vectors timelike; therefore, the distinction between
future lobes and past lobes cannot be univocally definable on a global
level. This means that the temporal orientability of space-time is a
precondition for defining a global arrow of time, since if space-time is not
temporally orientable, it is not possible to distinguish between the two
temporal directions for the universe as a whole.

However, not all accept this conclusion. For instance, Matthews \cite
{Matthews} claims that a space-time may have a regional but not a global
arrow of time if the arrow is defined by means of local considerations.
However, even from this local approach (which we have rejected in the
previous section), temporal orientability cannot be avoided. Let us suppose
that there were a local time-reversal non-invariant law $L$, which defines
regional arrows of time that disagree when compared by means of continuous
timelike transport. The trajectory of the transport will pass through a
frontier point between both regions: in a region around this point the arrow
of time will be not univocally defined, and this amounts to a breakdown of
the validity of $L$ in such a point. But this fact contradicts the
methodological principle of universality, unquestioningly accepted in
contemporary cosmology, according to which the laws of physics are valid in
all points of the space-time. The strategy to escape this conclusion would
consist in refusing to assign any meaning to the timelike continuous
transport. This strategy would only be acceptable if the two regions with
different arrows were physically isolated: this amounts to the
disconnectedness of the space-time. \ But this fact would contradict another
methodological principle of cosmology, that is, the principle of uniqueness,
according to which there is only one universe and completely disconnected
space-times are not allowed. These arguments show that the possibility of
time arrows pointing to opposite directions in different regions of the
space-time is not an alternative seriously considered in contemporary
cosmology.

Astronomical observations provide empirical evidence that makes implausible
the temporal non-orientability of our space-time. In particular, there is no
astronomical observation of temporally inverted behavior in some (eventually
very distant) region of the universe\footnote{%
In fact, supernovae evolutions always follow the same pattern (from
''birth'' to ''death''), and there is no trace of an inverted pattern in the
whole visible universe. This is a relevant fact when we consider that
supernovae are the markers used to measure the longest distances in our
universe, corresponding to objects near the observability horizon.}. On the
other hand, observational evidence in favor of the standard
Friedman-Lem\^{a}itre-Robertson-Walker models (FLRW, for short) plays the
role of indirect evidence for temporal orientability, since these
space-times are temporally orientable.

\subsection{Cosmic time}

As it is well known, general relativity replaces the older conception of
space-through-time by the concept of space-time, where time becomes a
dimension of a four-dimensional manifold. But when the time measured by a
physical clock is considered, each particle of the universe has its own {\it %
proper time}, that is, the time registered by a clock carried by the
particle. Since the curved space-time of general relativity can be
considered locally flat, it is possible to synchronize the clocks fixed to
particles whose parallel trajectories are confined in a small region of
space-time. But, in general, the synchronization of the clocks fixed to all
the particles of the universe is not possible. Only in certain particular
cases all the clocks can be coordinated by means of a cosmic time, which has
the features necessary to play the role of the temporal parameter in the
evolution of the universe.

The issue can also be posed in geometrical terms. A space-time may be such
that it is not possible to partition the set of all events into equivalent
classes such that: (i) each one of them is a spacelike hypersurface, and
(ii) the hypersurfaces can be ordered in time. There is a hierarchy of
conditions which, applied to a temporally orientable space-time, avoid
''anomalous'' temporal features (see \cite{Hawking-Ellis}). The strongest
condition is the existence of a global time.

{\bf Def.}: A {\it global time function} on the Riemannian manifold $M$ is a
function $t:M\rightarrow {\cal R}$ whose gradient is everywhere timelike.

In other words, the value of the global time function increases along every
future directed non-spacelike curve. The existence of such a function
guarantees that the space-time is globally splittable into hypersurfaces of
simultaneity which define a foliation of the space-time (see \cite{Schutz}).

Nevertheless, the fact that the space-time admits a global time function
does not yet permit to define a notion of simultaneity in an univocal manner
and with physical meaning. In order to avoid ambiguities in the notion of
simultaneity, we must choose a particular foliation. The foliation $\tau $
according to which all the worldline curves are orthogonal to all the
hypersurfaces $\tau =const.$ is the proper choice, because orthogonality
recovers the notion of simultaneity of special relativity for small regions
(tangent hyperplanes) of the hypersurfaces $\tau =const.$ (for the necessary
conditions see \cite{Misner}). However, even if this condition selects a
particular foliation, it permits that the proper time interval between two
hypersurfaces of simultaneity depends on the particular worldline considered
for computing it. If we want to avoid this situation, we must impose as an
additional constraint: the proper time interval between two hypersurfaces $%
\tau =\tau _{1}$ and $\tau =\tau _{2}$ must be the same for all worldline
curves. In this case, the metric results: 
\begin{equation}
ds^{2}=dt^{2}+h_{ij}\,dx^{i}\,dx^{j}  \label{1}
\end{equation}
where $t$ is the {\it cosmic time} and $h_{ij}=h_{ij}(t,x^{1},x^{2},x^{3})$
is the three-dimensional metric of each hypersurface of simultaneity.

Of course, the existence of a cosmic time imposes a significant topological
and metric limitation on the space-time. This means that, with no cosmic
time, there is not a single time which can be considered as the parameter of
the evolution of the universe and, therefore, it is nonsensical to speak of
the two directions of time for the universe as a whole. Therefore, the
possibility of defining a cosmic time is a precondition for meaningfully
speaking of a global arrow of time. This fact supplies an additional
argument against the entropic approach, which takes for granted the
possibility of defining the entropy function of the universe. But this
amounts to the assumption that: (i) the space-time can be partitioned in
spacelike hypersurfaces on which the entropy of the universe can be defined,
and (ii) the space-time possesses a cosmic time or, at least, a global time
on which the entropy gradient can be computed. When the possibility of
space-times with no cosmic time is recognized, it is difficult to deny the
conceptual priority of the geometrical structure of space-time over entropic
features in the context of our problem.

The question about the existence of a cosmic time has not a single answer
for all possible relativistic universes. But, what can we say about our
universe? Cosmology offers a simple answer on the basis of the cosmological
principle and the assumption of expansion. Since the universe is spatially
homogeneous and isotropic on the large scale, it is possible to find a
family of spacelike hypersurfaces which can be labeled by the proper time of
the worldlines that orthogonally thread through them: these labels define
the cosmic time. In the Robertson-Walker metric corresponding to flat FLRW
models: 
\[
ds^{2}=dt^{2}+a^{2}(t)\,\left( dx^{2}+dy^{2}+dz^{2}\right) 
\]
the cosmic time is represented by the variable $t$, and the scale factor $a$
is a scalar only function of $t$; this is the time by means of which
cosmologists estimate the age of the universe. In this sense, FLRW models
recover a notion of time analogous to the conception of pre-relativistic
physics, where time is an ordering parameter with respect to which the
evolution of the system is described.

\subsection{Time-asymmetry}

Of course, temporal orientability is merely a necessary condition for
defining the global arrow of time, but it does not provide a physical,
non-arbitrary criterion for distinguishing between the two directions of
time. As we will see, such a distinction requires the time-asymmetry of the
universe.

It is usually accepted that the obstacle for defining the arrow of time lies
in the fact that the fundamental laws of physics are time-reversal invariant%
\footnote{%
The exception is the law that rules weak interactions; but they are so weak
that it is difficult to see how the macroscopic arrow of time can be derived
from them. Therefore, as it is usual in the literature, we will not address
this question in this paper.}. Nevertheless, this common position can be
objected on the basis of the elucidation of the concepts of time-reversal
invariance and time-symmetry: whereas time-reversal invariance is a property
of dynamical equations (laws), time-symmetry is a property of a single
solution (evolution) of an dynamical equation.

{\bf Def.:} A solution $f(t)$ of a dynamical equation is {\it time-symmetric}
if there is a time $t_S$ such that $f(t+t_S)=f(t-t_S)$.

Therefore, the time-reversal invariance of an equation does not imply the
time-symmetry of its solutions: a time-reversal invariant law may be such
that all or most of the possible evolutions relative to it are individually
time-asymmetric. Huw Price \cite{Price} illustrates this point with the
familiar analogy of a factory which produces equal numbers of left-handed
and right-handed corkscrews: the production as a whole is completely
unbiased, but each individual corkscrew is asymmetric.

It is quite clear that these considerations are not applicable to the field
equations as originally stated. However, the existence of a cosmic time
allows to formulate the issue in familiar terms: under this condition,
Einstein's field equations are time-reversal invariant in the sense that if
the $h_{ij}(t,x^1,x^2,x^3)$ of eq.(1) is a solution, $h_{ij}(-t,x^1,x^2,x^3)$
is also a solution. But the time-reversal invariance of these equations does
not prevent us from describing a time-asymmetric universe whose space-time
is asymmetric regarding its geometrical properties along the cosmic time.
This idea can also be formulated in terms of the concept of time-isotropy.

{\bf Def.}: A temporally orientable space-time $(M,g)$ (where $M$ is a
four-dimensional pseudo-Riemannian manifold and $g$ is a Lorentzian metric
for $M$) is {\it time-isotropic} if there is a diffeomorphism $d$ of $M$
onto itself which reverses the temporal orientations but preserves the
metric $g$.

However, when we want to express the temporal symmetry of a space-time
having a cosmic time, it is necessary to strengthen the definition .

{\bf Def.}: A temporally orientable space-time which admits a cosmic time $t$
is {\it time-symmetric} with respect to some spacelike hypersurface $%
t=\alpha $, where $\alpha $ is a constant, if it is time-isotropic and the
diffeomorphism $d$ leaves fixed the hypersurface $t=\alpha $.

Intuitively this means that, from the hypersurface $t=\alpha $, the
space-time looks the same in both temporal directions. Therefore, if a
temporally orientable space-time having a cosmic time is time-asymmetric, we
will not find a spacelike hypersurface $t=\alpha $ which splits the
space-time in two ''halves'', one the temporal mirror image of the other
regarding their intrinsic geometrical properties.

When we turn our attention to the standard models of present-day cosmology,
we find that it is not difficult to apply these concepts. In FLRW models,
the time-symmetry of space-time may manifest itself in two different ways
according to whether the universe has singular points in one or in both
temporal extremities\footnote{%
This depends on the values of the factor $k$ and of the cosmological
constant $\Lambda $.}. Big Bang-Big Chill universes are manifestly
time-asymmetric: since the scale factor $a(t)$ increases with the cosmic
time $t$, there is no hypersurface $t=\alpha $ from which the space-time
looks the same in both temporal directions. In Big Bang-Big Crunch
universes, on the contrary, $a(t)$ has a maximum value: therefore, the
space-time might be time-symmetric about the time of maximum expansion: this
is the case of some FLRW models with dust and radiation. However, in more
general cases ({\it e.g.} inflationary models) it is necessary to add one or
many fields that represent the matter-energy of the universe. Many
interesting results have been obtained, for instance, by modeling
matter-energy as scalar fields $\phi _k(t)$: homogeneity is retained and the
time-reversal invariance of the field equations is given by the fact that,
if $\left[ a(t),\phi _k(t)\right] $ is a solution, $\left[ a(-t),\phi
_k(-t)\right] $ is also a solution. In these cases, if we call the time of
maximum expansion $t_{ME}$, the scale factor $a(t)$ may be such that $%
a(t_{ME}+t)\neq a(t_{ME}-t)$ (see, for instance, the models in \cite
{Cast-Giac-Lara}). This means that a Big Bang-Big Crunch universe may be a
time-asymmetric object regarding the metric of the space-time: this
asymmetry, essentially grounded on geometrical considerations, allows us to
distinguish between the two directions of the cosmic time, independently of
entropic considerations.

Up to this point we have argued for the possibility of describing
time-asymmetric universes by means of time-reversal invariant laws. But,
what is the reason to suppose that time-asymmetry has high probability? In
order to complete the argument, we will demonstrate that time-symmetric
universes are highly improbable to the extent that time-symmetric solutions
of the universe equations have measure zero in the corresponding phase space.

Let us consider some model of the universe equations. All known examples
have the following two properties (e.g. see \cite{Cast-Giac-Lara}, but there
are many other examples):

1.- They are time-reversal invariant, namely, invariant under the
transformation $t\rightarrow -t.$

2.- They are time-translation invariant, namely, invariant under the
transformation $t\rightarrow t+const.$\footnote{%
We are referring to the equations that rule the behavior of the universe,
not to the particular solutions that normally do not have time-translation
symmetry.} (homogeneous time).

Let us consider the generic case of a FLRW universe with radius $a$ and
matter represented by a neutral scalar field $\phi .$ The dynamical
variables are now $a,\stackrel{\bullet }{a},\phi ,\stackrel{\bullet }{\phi }$%
. They satisfy a generic Hamiltonian constraint\footnote{$H$ is the 00
component of Einstein equation (\ref{6.1}).}:

\begin{equation}
H(a,\stackrel{\bullet }{a},\phi ,\stackrel{\bullet }{\phi })=0  \label{4.0}
\end{equation}
which reduces the dimension of phase space from 4 to 3; then, we can
consider a phase space of variables $\stackrel{\bullet }{a},\phi ,\stackrel{%
\bullet }{\phi }$ and: 
\begin{equation}
a=f(\stackrel{\bullet }{a},\phi ,\stackrel{\bullet }{\phi })  \label{4.1}
\end{equation}
a function obtained solving eq.(\ref{4.0}).

If we want to obtain a time-symmetric continuous\footnote{%
We will disregard non-continuous solutions since normally information do not
pass through discontinuities and we are only considering {\it connected}
universes where information can go from a point to any other timelike
connected point.} solution such that $a\geq 0$\footnote{%
As only $a^2$ appears in a FLRW metric, we will consider just the case $%
a\geq 0$ since the point $a=0$ is actually a singularity that cuts the time
evolution.}, there must be a time $t_S$ regarding to which $a$ is symmetric:

\[
a(t_S+t)=a(t_S-t)\quad \text{ and\quad }\stackrel{\bullet }{a}(t_S)=0 
\]
In order to obtain complete time-symmetry, $\phi $ must also be symmetric
about $t_S$. There are two cases: even symmetry: 
\[
\phi (t_S+t)=\phi (t_S-t)\text{\quad and\quad }\stackrel{\bullet }{\phi }%
(t_S)=0 
\]
and odd symmetry: 
\[
\phi (t_S+t)=-\phi (t_S-t)\text{\quad and}\quad \phi (t_S)=0 
\]
This means that time-symmetric trajectories necessarily pass trough the axes 
$(0,\phi ,0)$ or ($0,0,\stackrel{\bullet }{\phi })$ of the phase space. From
these ''initial'' conditions we can propagate, using the evolution
equations, the corresponding trajectories; this operation will produce two
surfaces that contain the trajectories with at least one point of symmetry,
that is, that contain all the possible time-symmetric trajectories. Both
surfaces have dimension 2%
\mbox{$<$}%
3 (namely, the dimension of our phase space). The usual Liouville measure of
these sets is zero, and also any measure absolutely continuous with respect
to it. In this way we have proved that, for generic models of the universe,
the solutions are time-asymmetric with the exception of a subset of
solutions of measure zero. q.e.d.

This theorem can be easily generalized to the case where $\phi $ has many
components, or to the case of many fields with many components. Some of
these fields may be fluctuations of the metric: in this case, we must
Fourier transform the equations, and this would allow us to reproduce the
theorem only with $t$ functions. Since properties 1 and 2 (time-reversal
invariance and time-translation invariance) are also true in the classical
statistical case, the theorem can be also demonstrated in this case\footnote{%
When the phase space has infinite dimensions, it is better to use the notion
of dimension instead of that of measure.}. And also in the quantum case,
albeit some quantum gravity problems like time definition\cite
{Cast-Mazz-Lomb}.

\section{From the global arrow to the local arrow}

As we have seen, in a temporally orientable space-time a continuous
non-vanishing timelike vector field $\gamma ^\mu (x)$ can be defined all
over the manifold. At this stage, the universe is {\it temporally orientable}
but not yet {\it temporally oriented}, because the distinction between $%
\gamma ^\mu (x)$ and $-\gamma ^\mu (x)$ is just conventional. Now
time-asymmetry comes into play. In a temporally orientable time-asymmetric
space-time, any time $t_A$ splits the manifold into two sections that are
different to each other: the section $t>t_A$ is {\it substantially}
different than the section $t<t_A$. We can chose any point $x_0$ with $t=t_A$
and conventionally consider that $-\gamma ^\mu (x_0)$ points towards $t<t_A$
and $\gamma ^\mu (x_0)$ points towards $t>t_A$ or vice versa: in any case we
have established a substantial difference between $\gamma ^\mu (x_0)$ and $%
-\gamma ^\mu (x_0)$. We can conventionally call ''future'' the direction of $%
\gamma ^\mu (x_0)$ and ''past'' the direction of $-\gamma ^\mu (x_0)$ or
vice versa, but in any case past is substantially different than future. Now
we can extend this difference to the whole continuous fields $\gamma ^\mu
(x) $ and $-\gamma ^\mu (x)$: in this way, the time-orientation of the
space-time has been established. Since the field $\gamma ^\mu (x)$ is
defined all over the manifold, it can be used {\it locally }at each point $x$
to define the future and the past lobes: for instance, if we have called
''future'' the direction of $\gamma ^\mu (x)$, $C_{+}(x)$ contains $\gamma
^\mu (x)$ and $C_{-}(x)$ contains $-\gamma ^\mu (x)$.

Even if this solution is general for generic temporally orientable universes
having a cosmic time, it would be desirable to show how the global
time-orientation is reflected in everyday physics, where time-asymmetry
manifests itself in terms of time-asymmetric energy fluxes. This task will
lead us to impose reasonable restrictions in the considered cosmological
model in such a way that the explanation of local time-asymmetry applies,
not to the generic case, but rather to the particular case of our own
universe.

i.- Up to this point, global time-asymmetry has been considered as a
substantial asymmetry of the geometry of the universe, embodied in the
metric tensor $g_{\mu \nu }(x)$ defined at each point of the space-time.
Perhaps the easiest way to see how this geometrical time-asymmetry is
translated into local physical terms is to consider the energy-momentum
tensor $T_{\mu \nu }$, which can be computed by using $g_{\mu \nu }(x)$ and
its derivatives through Einstein's equation: 
\begin{equation}
T_{\mu \nu }-\frac{1}{8\pi }\,\left( R_{\mu \nu }(g)-\frac{1}{2}\,g_{\mu \nu
}\,R(g)-\Lambda \,g_{\mu \nu }\right) =0  \label{6.1}
\end{equation}
The curvatures $R_{\mu \nu }(g)$, $R(g)$ can be obtained from $g_{\mu \nu
}(x)$ and its derivatives, and $\Lambda $ is the cosmological constant. Now
we impose a first condition: that our $T_{\mu \nu }$ turns out to be a ''%
{\it normal}'' or {\it Type I} energy-momentum tensor. Then, $T_{\mu \nu }$
can be written as: 
\[
T_{\mu \nu }=s_{0}\,V_{\mu }^{(0)}\,V_{\nu
}^{(0)}+\sum_{i=1}^{3}\,s_{i}\,V_{\mu }^{(i)}\,V_{\nu }^{(i)} 
\]
where $\left\{ V_{\mu }^{(0)},\,V_{\mu }^{(i)}\right\} $ is an orthonormal
tetrad, $V_{\mu }^{(0)}$ is timelike and the $V_{\mu }^{(i)}$ are spacelike (%
$i=1,2,3$) (see \cite{Hawking-Ellis}, \cite{Lichne}). Since we have assumed
that the manifold is continuous, $g_{\mu \nu }(x)$ and also $T_{\mu \nu }(x)$
are continuously defined over the manifold (provided the derivatives of $%
g_{\mu \nu }(x)$ are also continuous); this means that $V_{\mu }^{(0)}(x)$
is a continuous unitary timelike vector field defined all over the manifold,
which can play the role of the field $\gamma ^{\mu }(x)$ if everywhere $%
s_{0}\neq 0$ (if not, $V_{\mu }^{(0)}$, even if timelike, may change its
sign when $s_{0}=0)$.

Here we impose a second condition: that the universe satisfies the {\it %
dominant energy condition: }i.e. $T^{00}\geq $ $\left| T^{\mu \nu }\right| $
in any orthonormal basis (namely, $s_0\geq 0$ and $s_i\in \left[
-s_0,s_0\right] $). In this case, $s_0\neq 0$ and, then, $V_\mu ^{(0)}(x)$
is continuous, timelike and non-vanishing. This means that $V_\mu ^{(0)}(x)$
can play the role of $\gamma ^\mu (x)$, with the advantage that it has a
relevant physical sense. In this way, a time-orientation is chosen at each
point $x$ of the manifold, and the time-components of $T_{\mu \nu }$ acquire
definite signs according to this orientation. Therefore we have translated
the global time-asymmetry into local terms, endowing the local arrow with a
physical sense.

ii.- Since we are now in local grounds, our new task is to understand the 
{\it local nature }of the characters in the play. If $T^{00}\geq $ $\left|
T^{\mu \nu }\right| $, then $T^{00}\geq $ $\left| T^{i0}\right| $.
Therefore, $T^{0\mu }$, which is usually but not rigorously conceived as the
local energy flux, can also be considered as the coordinates of a timelike
(or lightlike) vector. This holds for all presently known forms of
energy-matter and, so, there are in fact good reasons for believing that
this should be the case in almost all situations (for the exceptions, see 
\cite{Visser}\footnote{%
E.g., some exceptions are: Casimir effect, squeed vacuum, Hawking
evaporation, Hartle-Hawking vacuum, negative cosmological constant, etc.
These objects are strange enough in nowadays observational universe to
exclude the practical existence of zones with $T^{00}$ of different signs
and, therefore, with different time directions.}).

iii.-But, is really $T^{0\mu }$ the energy flux? To go even closer to
everyday physics, we must remember that $T_{\mu \nu }$ satisfy the
''conservation'' equation: 
\[
\nabla _{\mu }\,T^{\mu \nu }=0 
\]
Nevertheless, as it is well known, this is not a true conservation equation
since $\nabla _{\mu }$ is a covariant derivative. The usual conservation
equation with ordinary derivative reads: 
\[
\partial _{\mu }\,\tau ^{\mu \nu }=0 
\]
where $\tau _{\mu \nu }$ is not a tensor and it is defined as: 
\[
\tau _{\mu \nu }=\sqrt{-g}\ (T_{\mu \nu }+t_{\mu \nu }) 
\]
where we have introduced a $t_{\mu \nu }$ that reads: 
\[
\sqrt{-g}\ t_{\mu \nu }=\frac{1}{16\pi }\left[ {\cal L}\,g_{\mu \nu }-\frac{%
\partial {\cal L}}{\partial g_{\mu \nu },\lambda }\ g_{\mu \nu },\lambda
\right] 
\]
where ${\cal L}$ is the system\'{}s Lagrangian. $t_{\mu \nu }$ is also an
homogeneous and quadratic function of the connection $\Gamma _{\nu \mu
}^{\lambda }$ \cite{Landau}. Now we can consider the coordinates $\tau
^{0\mu }$, which satisfy: 
\[
\partial _{\mu }\tau ^{0\mu }=\partial _{0}\tau ^{00}+\partial _{i}\tau
^{0i}=0 
\]
namely, a usual conservation equation. Even if $\tau ^{0\mu }$ is not a
four-vector, it can be defined in each coordinate system: in each system $%
\tau ^{00}$ can be considered as the energy density and $\tau ^{0i}$ as the
energy flux (the Poynting vector). This means that the field $\tau ^{0\mu
}(x)$ represents the spatio-temporal energy flow within the universe better
than $T^{0\mu }$.

In particular, in a local inertial frame where $\Gamma _{\nu \mu }^\lambda
=0 $, we have $\tau _{\mu \nu }=\sqrt{-g}\ T_{\mu \nu }$: in orthonormal
coordinates, the dominant energy condition will be now $\tau ^{00}\geq $ $%
\left| \tau ^{i0}\right| $ and $\tau ^{0\mu }$ will be timelike (or
lightlike). But $\tau ^{0\mu }$ is just a local energy flow since it is
defined in orthonormal local inertial frames. Nevertheless, in any moving
frame with respect to the former one, if the acceleration of the moving
frame is not very large, the $(\Gamma _{\nu \mu }^\lambda )^2$ and the $%
t_{\mu \nu }$ are very small and the energy flux in the moving frame is
timelike (or lightlike) for all practical purposes. This is precisely the
case of the commoving frame of our present-day universe.

In summary, $\tau ^{0\mu }$ (that can locally be considered as the
four-velocity of a quantum of energy carrying a message) is a timelike {\it %
local} energy flux and:

a) It inherits the global time-asymmetry of $g_{\mu \nu }(x)$, i.e., the
geometrical time-asymmetry of the universe.

b) It translates the global time-asymmetry into the local level: the lobes $%
C_{-}(x)$ receive an incoming flux of energy while the lobes $C_{+}(x)$ emit
an outgoing flux of energy and, therefore, both kinds of lobes are
substantially different.

Thus we can consider that the energy flux $\tau ^{0\mu }$ is defined all
over the universe, and this local time-asymmetric flux is the agency that
produces time-asymmetry at every point within the universe. This phenomenon
has been explained in all details in papers \cite{TA}, where we have
introduced the classical Reichenbach-Davies diagram and the quantum
Reichenbach-B\"ohm diagram to illustrate it. In these contexts it is very
easy to deduce the different arrows of time (electromagnetic, quantum,
thermodynamic, etc.) from the global time-asymmetry of the universe. We
refer the reader to those papers to complete our view about the problem of
the arrow of time. In particular, in paper \cite{Olimpia} we have
established the substantial difference between t-symmetric twins
corresponding to several fields of physics. Here we will only add a new case
and make a relevant remark:

i.- In paper \cite{CCF} (Section III) we show that, in the Taub cosmological
model, the Hamiltonian can be written as: 
\[
H=\left( \sqrt{6\,}p_q+\frac 1{\sqrt{6}}\sqrt{\,p_u^2+(12\pi )^2\,e^{6u}}%
\right) \left( \sqrt{6\,}p_q-\frac 1{\sqrt{6}}\sqrt{\,p_u^2+(12\pi
)^2\,e^{6u}}\right) 
\]
showing a two sheet structure, that is, another case of t-symmetric twins
(clock-symmetric twins, in the language of paper \cite{CCF}). The constraint 
$H=0$ force us to choose one sheet-twin: the energy flux, which establishes
the substantial difference between the two members of the pair, supplies the
criterion for the selection.

ii.- The Postulate A.3 of the Axiomatic Quantum Field Theory (see \cite{Haag}%
, p.56, eq.II.1.15) sates that the spectrum of the energy-momentum operator $%
P^{\mu }$ is confined to the future light semi-cone, that is, its
eigenvalues $p^{\mu }$ satisfy: 
\[
p^{2}\geq 0\qquad \qquad p^{0}\geq 0 
\]
Condition $p^{0}\geq 0$ makes the theory (and, as a consequence, all
particle physics) a time-reversal non-invariant theory. But if we remember
that $\tau ^{0\mu }$ can be also considered as the linear momentum density, $%
p^{\mu }\sim \tau ^{0\mu }$, the condition $p^{0}\geq 0$ turns out to be a
consequence of $\tau ^{00}\geq $ $\left| \tau ^{i0}\right| $. Therefore,
instead of imposing Postulate A.3 as an axiom of the theory, it can be
justified on cosmological grounds.

\section{Conclusion}

The panorama is not completely closed yet: weak interactions should be
included in this scenario. However, this fact would not diminish the
relevance of the global non-entropic approach. From its very beginning,
theoretical physics has tried to combine its different chapters in an
unified formalism, and it is well known that unifications have always
produced great advances in physics. Therefore, our future challenge will be
to unify the weak-interactions explanation with the global explanation,
instead of abandoning the latter in favor of a local approach as many
local-minded physicists insist.

As it is well known, there is never a last word in physics. Nevertheless, we
can provisionally conclude that the global definition of the arrow of time
can be used as a solid basis for studying other problems related with the
time-asymmetry of the universe and its sub-systems.

\section{Acknowledgments}

We are very grateful to Gloria Dubner and Elsa Giacani for the footnote 1.
This paper was partially supported by CONICET and the University of Buenos
Aires.


\begin{references}
\bibitem{Irrev}  C. G. Sudarshan, C. B. Chiu and V. Gorini, {\it Phys. Rev. D%
}, {\bf 18, }2914, 1978. A. Bohm,{\it \ Quantum Mechanics, Foundations and
Applications}, Springer-Verlag, Berlin 1979. A. Bohm, M. Gadella, {\it Dirac
Kets, Gamow Vectors, and Gel'fand triplets}, Springer- Verlag, Berlin, 1989.
M. Castagnino and E. Gunzig, {\it Int. Jour. Theo. Phys}., {\bf 36}, 2545,
1997.

\bibitem{Penrose}  R. Penrose, ''Singularities and Time Asymmetry'', in S.
Hawking and W. Israel (eds.), {\it General Relativity, an Einstein Centenary
Survey}, Cambridge University Press, Cambridge, 1979.

\bibitem{Sachs}  R. G. Sachs, {\it The Physics of Time-Reversal}, University
of Chicago Press, Chicago, 1987.

\bibitem{Price}  H. Price, {\it Time's Arrow and the Archimedes' Point, }%
Oxford University Press, Oxford, 1996.

\bibitem{Ehrenfest}  P. Ehrenfest and T. Ehrenfest, {\it The Conceptual
Foundations of the Statistical Approach in Mechanics}, Cornell University
Press, Ithaca, 1959 (original 1912).

\bibitem{Lax-Phillips}  P. D. Lax and R. S. Phillips,{\it \ Scattering Theory%
}, Academic Press, New York, 1979.

\bibitem{Cast-Laura}  M. Castagnino and R. Laura, {\it Phys. Rev. A}, {\bf 56%
}, 108, 1997.

\bibitem{Boltzmann}  L. Boltzmann, {\it Ann. Phys}., {\bf 60, }392, 1897.

\bibitem{Feynman}  R. P. Feynman, R. B. Leighton and M. Sands, {\it The
Feynman Lectures on Physics, Vol. 1, }G. Mattews, {\it Phil. Scie}., {\bf %
46, }82, 1979.Addison-Wesley{\it , }New York, 1964.

\bibitem{Davies}  P. C. Davies, ''Stirring Up Trouble'', in J. J. Halliwell,
J. Perez-Mercader and W. H. Zurek (eds.), {\it Physical Origins of Time
Asymmetry}, Cambridge University Press, Cambridge, 1994.

\bibitem{Earman}  J. Earman, {\it Phil. Scie}., {\bf 41, }15, 1974.

\bibitem{Matthews}  G. Mattews, {\it Phil. Scie}., {\bf 46, }82, 1979.

\bibitem{Hawking-Ellis}  S. Hawking and J. Ellis, {\it The Large Scale
Structure of Space-Time, }Cambridge University Press, Cambridge, 1973.

\bibitem{Schutz}  B. F. Schutz, {\it Geometrical Methods of Mathematical
Physics}, Cambridge University Press, Cambridge, 1980.

\bibitem{Misner}  C. W. Misner, K. S. Thorne and J. A Wheeler, {\it %
Gravitation}, Freeman and Co., San Francisco, 1973.

\bibitem{Cast-Giac-Lara}  M. Castagnino, H. Giacomini and L. Lara, {\it %
Phys. Rev. D}, {\bf 61}, 107302, 2000; {\bf 63}, 044003, 2001.

\bibitem{Cast-Mazz-Lomb}  M. Castagnino, {\it Phys. Rev. D}, {\bf 29, }2216,
1989. M. Castagnino and F. D. Mazzitelli, {\it Phys. Rev. D}, {\bf 42, }482,
1990. M. Castagnino and F. Lombardo, {\it Phys. Rev. D}, {\bf 48, }1722,
1993.

\bibitem{Lichne}  A. Lichnerowicz, {\it Th\'{e}ories Relativistes de la
Gravitation et de l'Electromagn\'{e}tisme}, Masson, Paris, 1955.

\bibitem{Visser}  M. Visser, {\it Lorentzian Wormholes}, Springer-Verlag,
Berlin, 1996.

\bibitem{Landau}  L. Landau and E. Lifchitz, {\it Th\'{e}orie des Champs},
Editions Mir, Moscow, 1970.

\bibitem{TA}  M. Castagnino, F. Gaioli and E. Gunzig, {\it Found. Cosmic
Phys.}, {\bf 16, }221, 1996. M. Castagnino and E. Gunzig, {\it Int. Jour.
Theo. Phys}., {\bf 36}, 2545, 1997. M. Castagnino, ''The global nature of
time asymmetry and the Bohm-Reichenbach diagram''{\it , }in A. Bohm, H.
Doebner and P. Kielarnowski (eds), {\it Irreversibility and Causality (Proc.
G. 21 Goslar 1996)}, Springer-Verlag, Berlin, 1998. M. Castagnino and E.
Gunzig, {\it Int. Journ. Theo. Phys}. {\bf 38,} 47, 1999. M. Castagnino, J.
Gueron and A. Ordo\~{n}ez, {\it J. Math. Phys}., {\bf 43}, 705, 2002. M.
Castagnino and C. Laciana, {\it Class. Quant. Grav.}, {\bf 19,} 2657, 2002.

\bibitem{Olimpia}  M. Castagnino, L. Lara and O. Lombardi, ''The
cosmological origin of time-asymmetry'', submitted to {\it Class. Quant.
Grav.}, LANL quant-ph/0211163.

\bibitem{CCF}  M. Castagnino, G. Catren and R. Ferraro, ''Time asymmetries
in quantum cosmology and the searching for boundary conditions to the
Wheeler-De Witt equation'', {\it Class. Quant. Grav}., in press, 2002.

\bibitem{Haag}  R. Haag, {\it Local Quantum Physics. Fields, Particles,
Algebras}, Springer, Berlin, 1996.
\end{references}
\end{document}